# Wafer-scale uniformity of Dolan-bridge and bridgeless Manhattan-style Josephson junctions for superconducting quantum processors

Nandini Muthusubramanian[1,2]‡, Pim Duivestein[1,2]‡, Christos Zachariadis[1,2]§, Matvey Finkel[1,2], Sean L. M. van der Meer[1,2], Hendrik M. Veen[1,2], Marc W. Beekman[1,3], Thijs Stavenga[1,2], Alessandro Bruno[1,2]§, and Leonardo DiCarlo[1,2]

[1] QuTech, Delft University of Technology, P.O. Box 5046, 2600 GA Delft, The Netherlands

[2] Kavli Institute of Nanoscience, Delft University of Technology, P.O. Box 5046, 2600 GA Delft, The Netherlands

[3] Netherlands Organisation for Applied Scientific Research (TNO), P.O. Box 155, 2600 AD Delft, The Netherlands

‡ Present address: QphoX B.V., Elektronicaweg 10, 2628 XG Delft, The Netherlands
§ Present address: Quantware B.V., Elektronicaweg 10, 2628 XG Delft, The Netherlands

## 1. Abstract

We investigate die-level and wafer-scale uniformity of Dolan-bridge and bridgeless Manhattan Josephson junctions, using multiple substrates with and without through-silicon vias (TSVs). Dolan junctions fabricated on planar substrates have the highest yield and lowest room-temperature conductance spread, equivalent to $\sim$ 100 MHz in transmon frequency. In TSV-integrated substrates, Dolan junctions suffer most in both yield and disorder, making Manhattan junctions preferable. Manhattan junctions show pronounced conductance decrease from wafer centre to edge, which we qualitatively capture using a geometric model of spatially-dependent resist shadowing during junction electrode evaporation. Analysis of actual junction overlap areas using scanning electron micrographs supports the model, and further points to a remnant spatial dependence possibly due to contact resistance.

## 2. Introduction

Monolithic superconducting quantum processors (SQPs) have scaled to enable key demonstrations of quantum-computational advantage [1] and milestone demonstrations of quantum error correction [2, 3, 4] on the road to fault-tolerant quantum computing. Sustaining this scaling requires a multi-faceted fabrication approach simultaneously meeting yield, frequency, coherence, and coupling requirements of circuit elements, as well as the routing of control lines needed for gate and measurement operations. The latter motivates the active development of 3D integration strategies such as flip-chip [5, 6, 7] to avoid overcrowding of circuit elements and vertical routing [8, 9, 10] of input and output lines to circumvent the scaling limitations associated with lateral wirebonding. Through-silicon visas (TSVs) are needed in some vertical routing approaches [11, 12, 13, 14], and especially for suppression of resonance modes arising from the increased size of SQPs and their packaging.

TSVs further aggravate the targeting of superconducting qubit frequencies, which already bottlenecks the yield of operable devices even on planar substrates [15]. Poor qubit frequency targeting is a primary cause of crosstalk induced by microwave drives [2] and can limit gate speeds. It also increases residual $ZZ$ coupling in processors with always-on qubit-qubit coupling [16, 2, 3], making gate fidelity and leakage dependent on the state of spectator qubits [17]. Laser annealing of qubit Josephson junctions (JJs) [18, 15, 19, 20, 21] is an established method for selective qubit frequency trimming without intrinsic effect on qubit coherence. Currently, laser annealing allows a monotonic decrease with $\sim$ 300 MHz range and $\sim$ 15 MHz precision. To safely rely on laser annealing for post-fabrication trimming, fabrication itself must target qubit frequencies with a precision of $\sim$ 50 MHz.

The main limit to qubit frequency targeting is variability in the fabrication of Al-AlO$_x$-Al JJs, which most commonly relies on double-angle shadow evaporation with intermediate in-situ oxidation. Two main variables affecting the Josephson coupling energy are the overlap area between the two Al electrodes and the tunnel barrier thickness. The two most popular JJ fabrication variants differ only in the shadowing mechanism: Dolan [22] JJs use a suspended resist bridge whereas Manhattan [23] junctions do not. Since Dolan JJs are more sensitive to resist-height variation by design, Manhattan junctions may be preferable particularly on substrates with TSVs that compromise the uniformity of spin-coated resist. On the other hand, recent reports by colleagues and us [24, 25, 26] indicate that geometric effects cause pronounced centre-to-edge variation in Manhattan JJs, affecting their uniformity at wafer scale.

In this work, we present an experimental investigation comparing the uniformity of Dolan versus Manhattan JJs at both die- and wafer-scale on planar substrates with and without TSVs. We benchmark uniformity using room-temperature (RT) conductance ($G$) measurements, extracting the conductance coefficient of variation (CV) and residual standard deviation (RSD) of predicted transmon frequency. A wafer-centre-to-edge variation is again observed particularly in Manhattan junctions, which we attribute to a geometric shadowing effect during electrode evaporation. Scanning electron microscopy (SEM) of many junctions supports the model, and further points to remnant spatial dependence possibly due to contact resistance. Our findings indicate that for our current fabrication processes, Dolan JJs perform best for planar substrates, while the opposite holds for TSV-integrated substrates. We identify several paths for further required improvement.

## 3. Design of experiments

We investigate uniformity of Dolan and Manhattan JJs using six 100-mm diameter Si wafers. (Sections S1 and S2 of the Supplementary Information provide detailed descriptions of the fabrication processes used.) Three of these wafers, labelled Planar 17Q (quantity one) and TSV 17Q (quantity two), are used to obtain and compare metrics for both junction variants in fully planar substrates and TSV-integrated ones. Each wafer contains thousands of test structures, each consisting of two nominally identical JJs connecting in parallel to pre-fabricated NbTiN probing pads (200 nm thick, defined by sputtering and etching). These test structures mimic the two-junction transmon with NbTiN capacitor plates used in our standard SQPs (figure 1(a)).

In the Planar 17Q wafer, a $13 \times 13$ mm die-level layout mimicking our planar 17-qubit SQP (Surface-17 [11, 27, 21]) is copy-pasted into two $2\times 4$ arrays, the top (bottom) array with Dolan (Manhattan) test structures. At the location of each transmon of the SQP, we place a sub-array of $4 \times 4$ test structures. Within each sub-array (figure 1(b)),

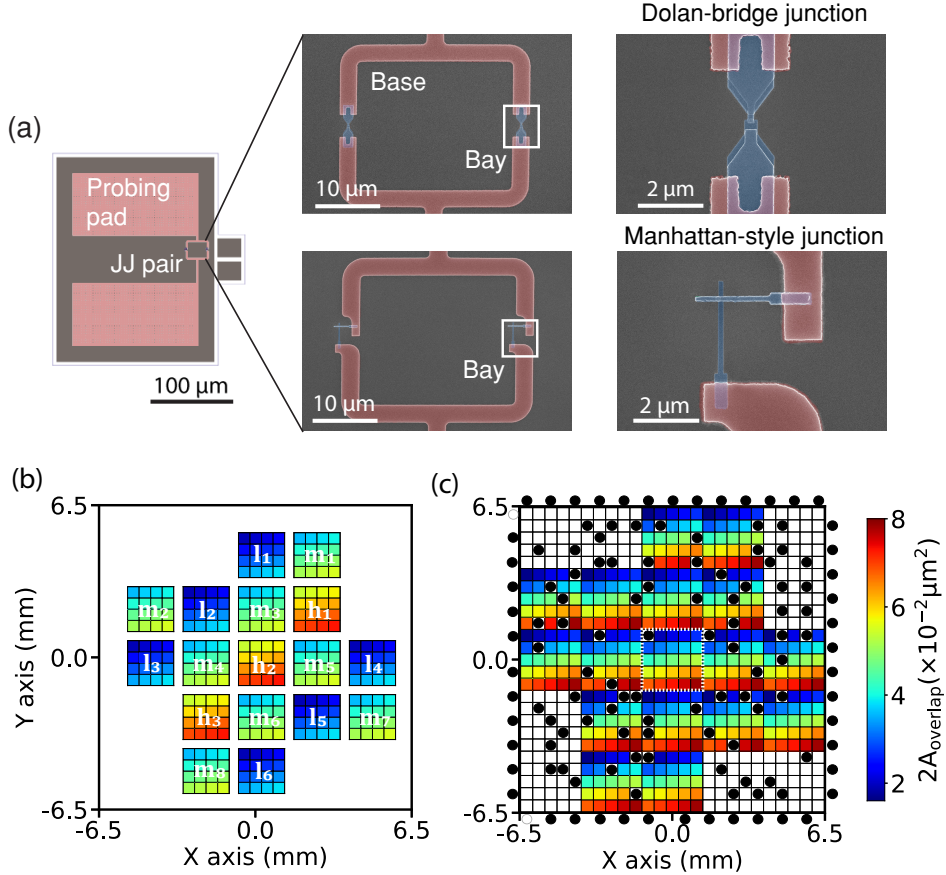

Figure 1: (a) Schematic and SEM images at two length scales of the test structures used to investigate uniformity of Dolan versus Manhattan JJ pairs on planar and TSV-integrated wafers. Two junctions in each structure complete a loop with a pre-fabricated NbTiN base. Probing pads in the base allow measuring the parallel conductance of the junction pair. (b) Die-level planar layout with 17, $4 \times 4$ sub-arrays of junction test structures. Each array is centred at the location of one transmon in our planar Surface-17 SQP. Each array has a sweep of junction overlap area $A_{\text{overlap}}$ in one of three ranges (labelled l, m and h). (c) Die-level TSV layout arranged as 17, $5 \times 5$ sub-arrays of junction test structures. Each array is centred at the location of one transmon in our TSV-integrated Surface-17. One such array is highlighted by the white dotted line. Each array has an identical sweep of $A_{\text{overlap}}$. Test structures that overlap with vias (black circles) are ignored and not included in measurements, yielding at most 378 test structures per die. Heatmaps in (b) and (c) indicate the chosen $A_{\text{overlap}}$ for each test structure.

the designed single-junction overlap area ($A_{\text{overlap}}$) is finely stepped within one of three ranges, labelled low (l), mid (m) and high (h), mimicking the choice of three qubit-frequency groups in our SQPs [11, 16, 27, 21]. For Dolan structures, we change $A_{\text{overlap}}$ by varying the width $W_{\text{t}}$ of the top electrode and keeping the width of the bottom electrode $W_{\text{b}} = 3W_{\text{t}}$. For Manhattan structures, we instead vary $W_{\text{b}}$ and fix

$W_\mathrm{t} = 160$ nm. In total, the wafer contains 2176 test structures of each JJ variant.

Each TSV 17Q wafer contains test structures of only one JJ variant. In each wafer, the die-level layout (copy-pasted into one $2 \times 4$ array) has TSVs placed at the same locations as a variant of Surface-17 with TSVs (figure 1(b)). The density ($\sim 1.7\%$ area coverage) and position of TSVs is chosen to push the lowest-frequency spurious modes of the SQP in its sample holder to $\gtrsim 15$ GHz (as per finite-element simulation). At the location of each transmon in the SQP, we place a $5 \times 5$ sub-array of test structures. In this case, all sub-arrays are identical. Importantly, test structures overlapping with TSVs, although fabricated, are ignored and not included in conductance measurements. This yields at most 378 viable test structures per die and thus 3024 per wafer.

Three additional wafers, labelled Planar $35 \times 35$, are used to test the geometric resist-shadowing model and to investigate further sources of spatial non-uniformity in Manhattan JJs. Each wafer (figure 5) has a $35 \times 35$ array of nominally identical test structures ($W_\mathrm{b} = W_\mathrm{t} = 200$ nm). In the first wafer, like in the 17Q wafers, the test structures have symmetric JJ pairs with NbTiN probing pads. In the second, they have symmetric JJ pairs with TiN probing pads (160 nm thick, pre-defined by atomic layer deposition (ALD) and etching). In the third, they have single JJs with Al probing pads evaporated simultaneously with the JJ electrodes.

## 4. Measurements and analysis

All $G$ measurements are acquired by the 2-point method using a home-built transimpedance amplifier. A low input voltage (10 mV) is applied across the junctions to minimize the possibility of causing failure to open or short circuit. Measurements on all planar wafers are performed using a manual probe station, with one exception noted below. During manual measurements, the intensity from a light-emitting diode source is set to the lowest possible visibility ($< 500$ lx) to minimize the parallel conductance contribution from the Si substrate to $\sim 5$ µS. Measurements on the TSV 17Q wafers as well as on the Planar $35 \times 35$ TiN wafer are performed using a home-built automated probe station. The measured series resistance contribution from external cabling is $< 10$ Ω. The series resistance of NbTiN probing pads was found to vary from 200 Ω at wafer centre to 330 Ω at wafer edge by fabricating test structures with bays short-circuited directly in the base layer. This variation is attributed to the radial dependence of the thickness of the sputtered NbTiN films (resistivity $\rho = 100$ µΩ-cm).

The range of $G$ is $40-350$ µS. Values $< 20$ µS and $> 500$ µS are filtered out as they mostly correspond to open and shorted junctions, respectively. To systematically detect and filter out data containing an open junction in a pair, a two-part linear regression analysis of conductance versus $A_\mathrm{overlap}$ is implemented within each die in the Planar and TSV 17Q wafers. Values below 70% of the initial best fit are filtered out (figure S2 and S3). For the Planar 35 × 35 wafers containing nominally identical test structures throughout, conductance values below 70% of the mean are filtered out.

To quantify non-uniformity at both die and wafer scale, we use the conductance CV as a function of $A_{\text{overlap}}$ and the RSD of predicted qubit frequency. Die-(wafer-)level conductance CV is calculated using all the test structures with identical $A_{\text{overlap}}$ across the die (wafer) when calculating the mean $\mu_G$ and standard deviation $\sigma_G$. The spatial variation of junction conductance is visualized using heatmaps of conductance normalized by $\mu_G$ of all test structures with identical $A_{\text{overlap}}$. The predicted transmon qubit transition frequency ($f_{01}$) is calculated from $G$ using

$$f_{01} = \sqrt{8f_{\text{C}}MG} - f_{\text{C}},$$

where $f_{\text{C}} = E_{\text{C}}/h = 270$ MHz is the designed transmon charging energy and $M = 134$ GHz/mS is an experimentally determined constant [28, 29, 30]. Die-level frequency RSD is calculated from the residuals of the second fit. Wafer-level RSD is calculated similarly, but the residuals are obtained by performing a single fit on the combined filtered $G$ data from all dies.

To test the geometric resist-shadowing model, SEM images of JJs from different coordinates on all Planar 35 × 35 wafers are acquired at $10^5\times$ magnification. SEM imaging is only performed after conductance measurements are completed. The actual deposited junction widths ($W'_{\text{b}}$, $W'_{\text{t}}$) and overlap area ($A'_{\text{overlap}}$) are extracted using home-made image analysis software (based on the OpenCV package) with the work flow presented in figure S7. The presence of other sources of spatial non-uniformity is evidenced from the spatial dependence of effective JJ conductivity calculated as $G/\Sigma A'_{\text{overlap}}$.

## 5. Results

A total of 2176 (3024) test structures are fabricated per JJ variant for the Planar and TSV 17Q datasets. A zoomed-out view (figure 2) of the planar dataset shows that the spatial variation of normalized conductance for Dolan JJs is significantly lower than for Manhattan JJs. For the latter, there is a clear systematic decrease from centre to edge, making it unsurprising that the wafer-scale conductance CV is higher for Manhattan over all $A_{\text{overlap}}$. The general decrease observed in the conductance CV with increasing $A_{\text{overlap}}$ is in line with previous works [31, 32]. At the die level, the spread of Dolan JJs is also lowest, with $\sim 100$ MHz frequency RSD uniform across the wafer. For Manhattan JJs, the frequency RSD increases away from wafer centre, indicating that the spatial variation is relevant even at die level.

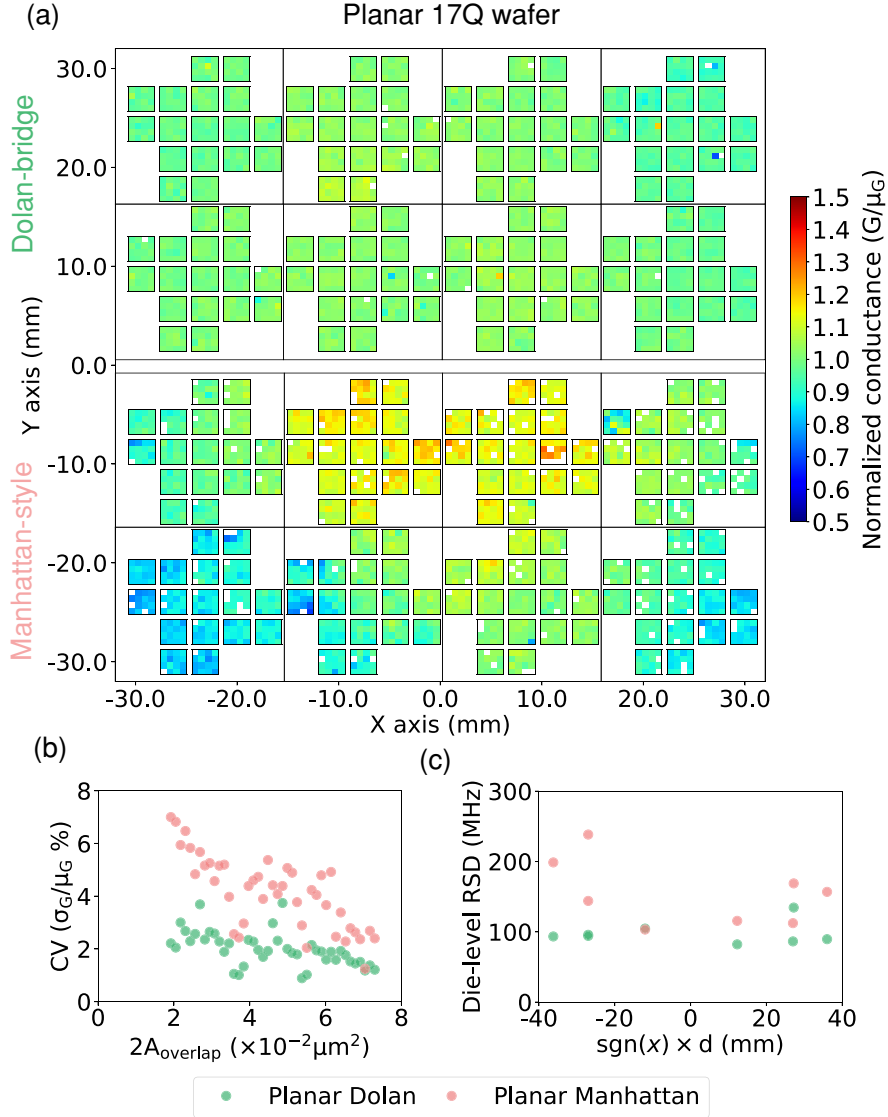

Figure 2: (a) Wafer-scale mean-normalized conductance heatmap of Dolan (top) and Manhattan (bottom) JJ test structures on the Planar 17Q wafer. The origin $(0,0)$ indicates wafer centre. Blank cells correspond to test structures identified as defective by the filtering. For this dataset, both JJ types are fabricated on a single wafer. (b) Wafer-scale conductance CV for both junction types as a function of $A_{\text{overlap}}$. (c) Die-level RSD of predicted qubit frequency as a function of distance ($d$) between die and wafer centres.

Turning over to the TSV dataset (figure 3), we can again discern an underlying centre-to-edge dependence for Manhattan JJs. However, this trend is masked by a significant increase in disorder. The disorder is much stronger for Dolan JJs, evident both at wafer scale and die level. Interestingly, the CV for Dolan does not display any clear dependence on $A_{\text{overlap}}$, suggesting that resist-height variations dominate the spread. Note that the CV and RSD for Dolan are calculated both with and without applying regression filters. This is necessary because the high disorder makes the regression filter unable to reject only defective junctions. Even with the artificial improvement of Dolan CV and RSD that may arise from removing non-defective junctions, a strong conclusion holds: with TSVs, Manhattan JJs systematically outperform Dolan JJs. Nonetheless, with > 300 MHz RSD at die level, even Manhattan JJs fall very short of frequency targeting objectives in the presence of TSVs. However, there is room for optimism as this investigation is best interpreted as a worst-case scenario for actual TSV-integrated SQPs. In our test, we place many junction pairs per transmon location of Surface-17. Therefore, in a real Surface-17, transmon JJ pairs would on average be farther away from TSVs. Furthermore, the footprint of TSVs could be further optimized following [12].

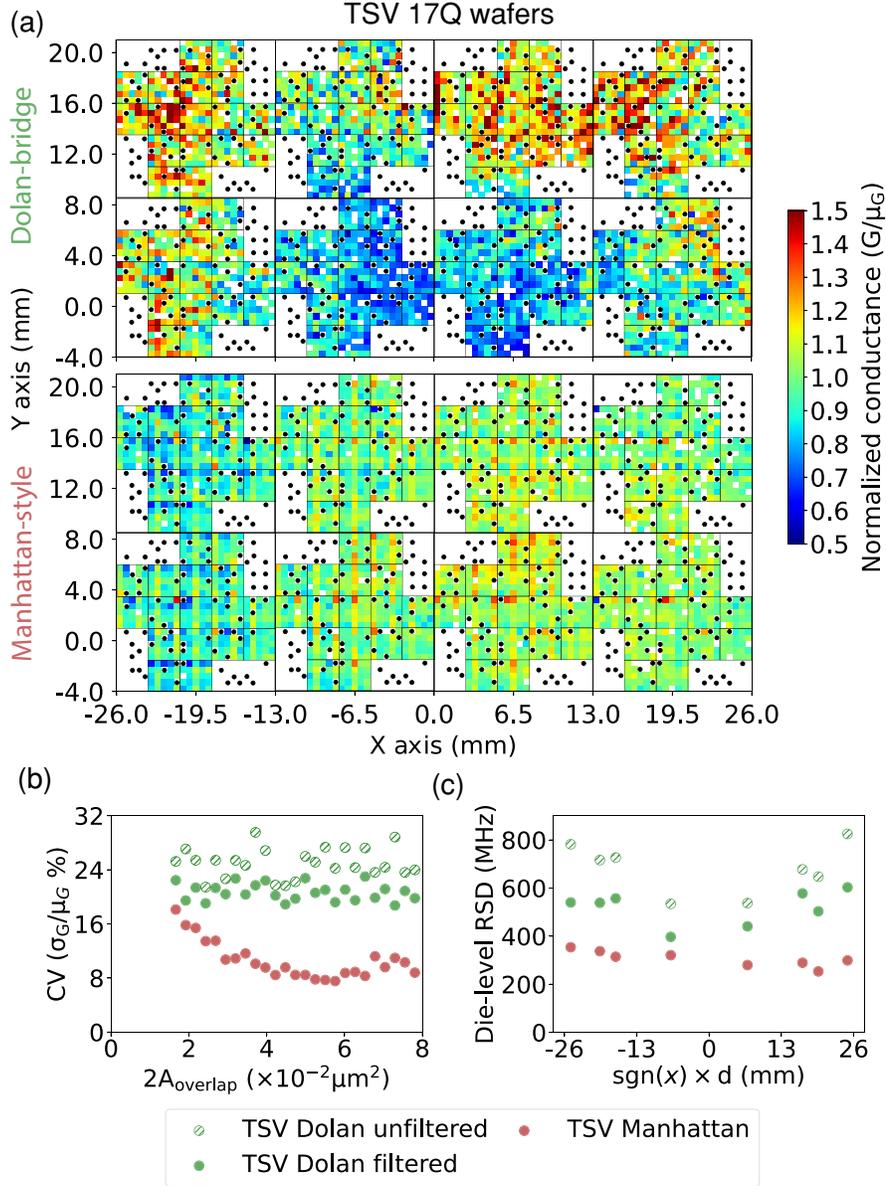

Figure 3: (a) Wafer-scale mean-normalized conductance heatmap of Dolan (top) and Manhattan (bottom) JJ test structures on TSV-integrated 17Q wafers. For this dataset, two separate wafers are fabricated, one for each JJ type. The origin $(0,0)$ indicates wafer centre. Blank cells correspond to defective junctions removed by filtering outliers at die level. Cells marked with black circles indicate TSV locations. (b) Wafer-scale conductance CV for unfiltered (nf) and regression-filtered (f) Dolan JJ pairs and for filtered Manhattan JJ pairs as a function of $A_{\text{overlap}}$. (c) Die-level RSD of predicted qubit frequency as a function of distance ($d$) between die and wafer centres.

## Summary of results

| Junction type | Substrate | Yield (%) | Conductance CV wafer scale (%) | Frequency RSD wafer scale (MHz) | Frequency RSD die level (MHz) |
|---|---|---|---|---|---|
| Dolan-bridge | Planar 17Q NbTiN | 2160/2176 = 99.2 | 0.8-3.7 | 140 | 98 |
|  | TSV 17Q NbTiN | 2958/3024 = 97.8 | 21.6-29.5 | 800 | 681 [a] |
|  |  | 2770/3024 = 91.6 | 18.5-22.5 | 666 | 520 [b] |
| Manhattan-style | Planar 17Q NbTiN | 2006/2176 = 92.2 | 1.2-7 | 317 | 155 |
|  | TSV 17Q NbTiN | 2867/3024 = 94.8 | 6.5-17 | 342 | 306 |
|  | Planar 35 × 35 NbTiN | 1176/1225 = 96.0 | 11.3 | - | - |
|  | Planar 35 × 35 TiN | 1161/1225 = 94.8 | 8.9 | - | - |
|  | Planar 35 × 35 Al | 1121/1155 = 97.0 [c] | 6.8 | - | - |

[a] Without regression filtering.
[b] With regression filtering.
[c] Two rows were accidentally omitted during data acquisition.

Table 1: Summary of metrics obtained for Dolan and Manhattan JJ test structures on all wafers used throughout this study. The die-level frequency RSD is the average across the eight dies in the 17Q wafers.

### 5.1. Geometric resist-shadowing model

The essence of the geometric model is a spatial dependence of junction electrode widths arising from oblique incidence of the Al flux during evaporation. Specifically, the width of vertical electrodes (both electrodes for Dolan JJs, bottom electrode for Manhattan JJs) depends on the $x$ coordinate, while that of horizontal electrodes (top electrode for Manhattan) depends on the $y$ coordinate. Key parameters of the model are the thickness of the top resist $H = 600$ nm (which acts as the shadow mask), the wafer tilt $\alpha = 35°$ during Al evaporations, and the physical configuration of the electron-beam (e-beam) evaporator (Plassys MEB550S). These last parameters include the distance $D' = 650$ mm between the crucible at $\vec{C}$ and the pivot point $\vec{O}'$ of the sample holder, and the distance $R = 62.5$ mm between $\vec{O}'$ and centre $\vec{O}$ of the exposed wafer surface (see schematic in figure 4(a)). This results in a distance $D = D' \cos(\alpha) - R$ between $\vec{C}$ and the plane defined by this surface [33]. In a cartesian coordinate system with origin at $\vec{O}$ and $\vec{r} = (x, y, 0)$ lying on this plane, $\vec{C} = (0, D'\sin(\alpha), D)$. Evaporation under these conditions deposits electrodes extending along the $y$ axis. An electrode of this orientation with $x$ coordinate has actual width

$$W'(x) \approx W + \delta W_{\text{offset}} - |x|\frac{H}{D}, \tag{1}$$

where $\delta W_{\text{offset}}$ is a constant widening from over-exposure and development of the e-beam resist. Including these modifications to the width of both electrodes, the actual overlap area becomes

$$A'_{\text{overlap}}(\vec{r}) = W'_{\text{b}}(x) W'_{\text{t}}(y). \tag{2}$$

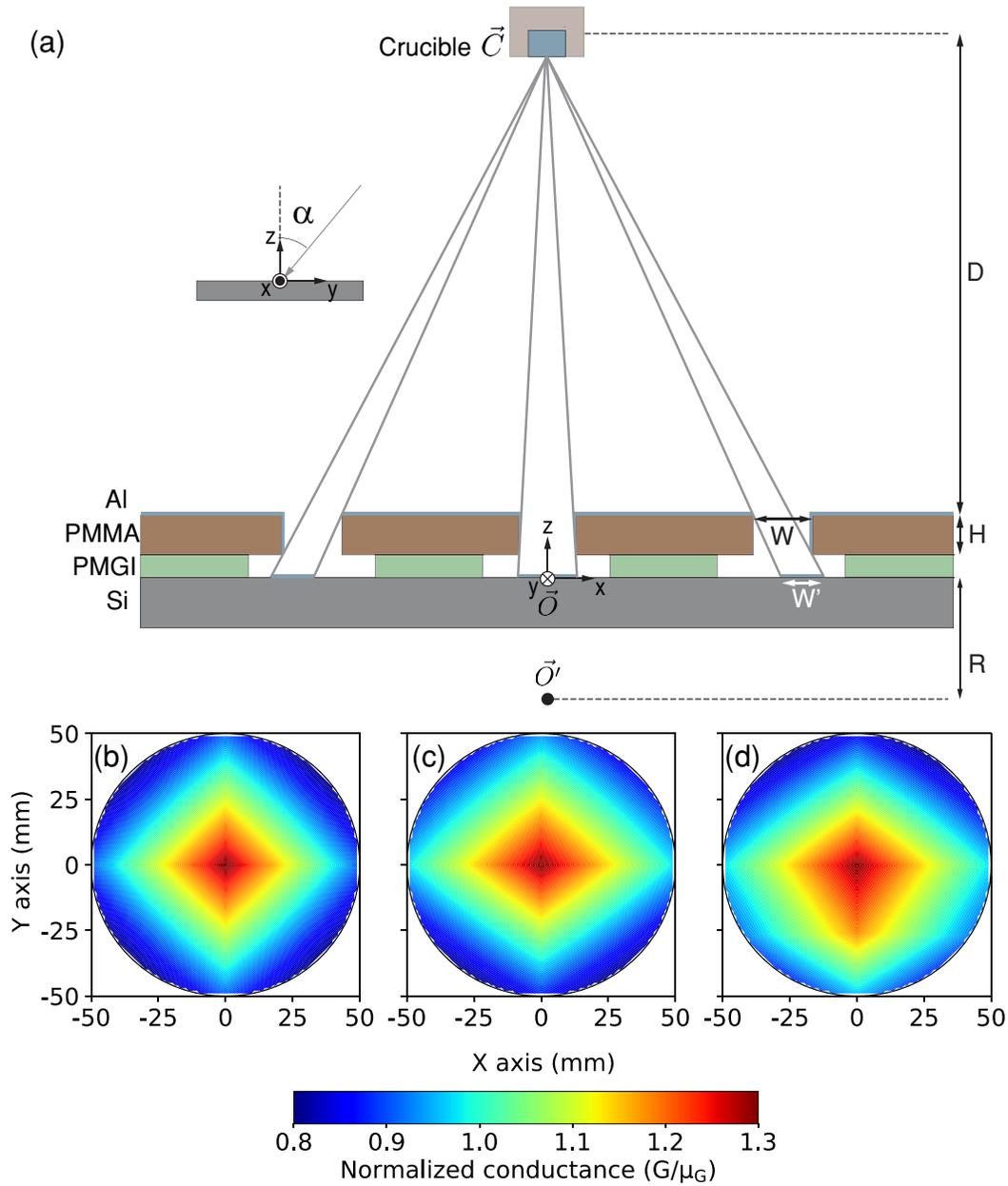

Figure 4: (a) Schematic of e-beam Al evaporation setup (not drawn to scale). Please see text for further details and parameter values. The illustration shows the decrease in junction electrode width from centre to edge of wafer arising from the spatially-dependent shadowing effect. (b) Wafer-scale mean-normalized conductance computed from actual junction overlap area $A'_{\text{overlap}}$ as per equation 2, for Manhattan JJs with $W_{\text{t}} = W_{\text{b}} = 200$ nm and $\delta W_{\text{offset}} = 25$ nm. (c) Same as (b) but adding the overlap contribution from sidewalls as per equation 4. (d) Same as (c) but adding effects of the first evaporation (of the bottom electrode) on the second evaporation (of the top electrode) (equations 5 to 7).

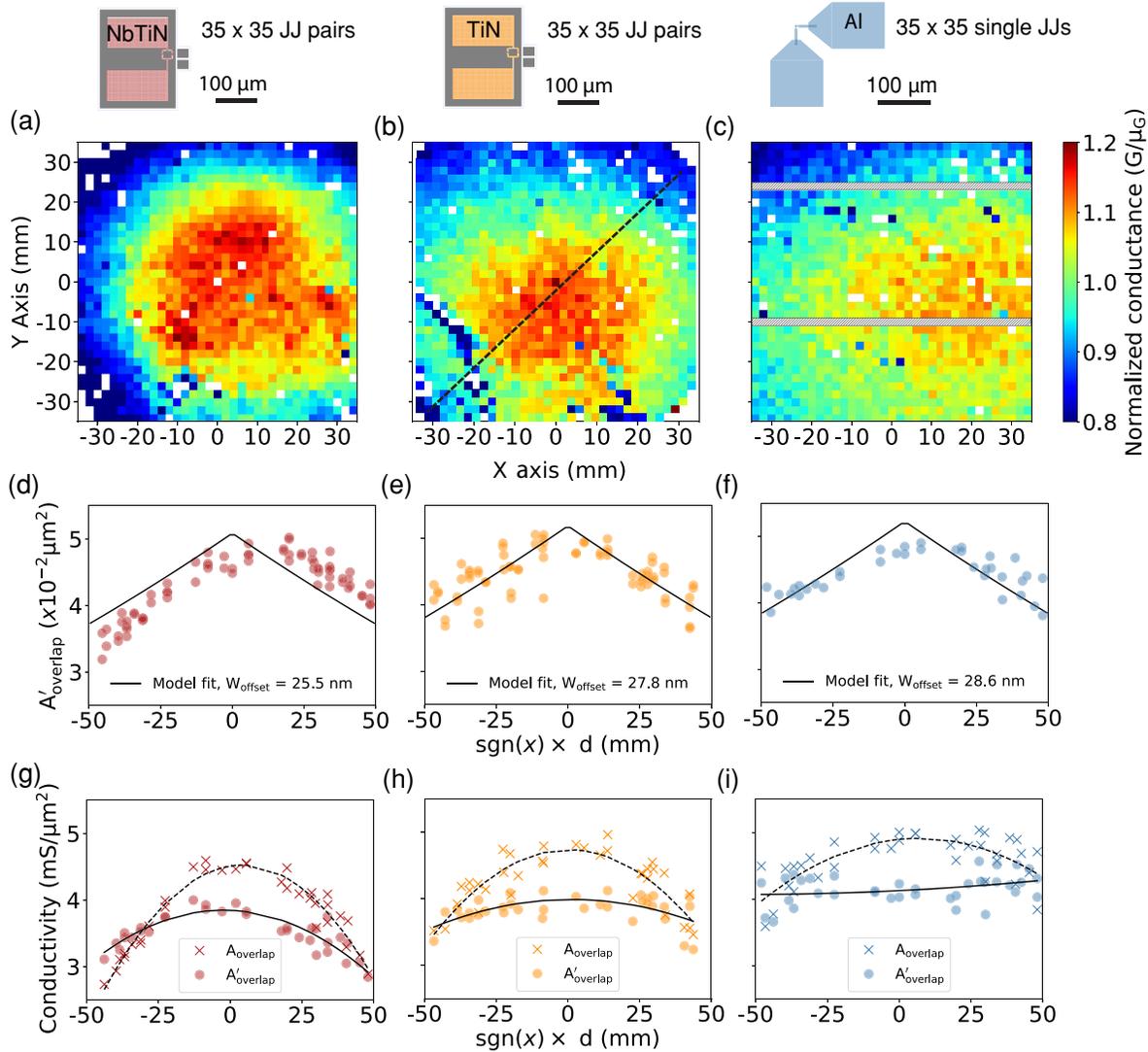

Figure 5: Wafer-scale mean-normalized conductance heatmap of $35 \times 35$ array of Manhattan JJ test structures fabricated on three planar wafers with the variants indicated by the top schematics. (a,b) Symmetric junction pairs with (a) NbTiN probing pads deposited by sputtering and (b) TiN probing pads deposited by ALD. The black dotted line indicates the diagonal along which the JJ pairs are imaged for figure S8. (c) Single junctions with simultaneously fabricated Al probing pads. The hatched rows indicate accidentally omitted junctions during data acquisition. (d–f) Distribution of actual junction overlap area $A'_{\text{overlap}}$ as a function of junction radial position ($d$). The black curves are the best fits of the simplest geometric model (equation 2 with single free parameter $\delta W_{\text{offset}}$). (g–i) Effective junction conductivity (computed from designed and actual overlap areas) as a function of $d$. The dashed (solid) curves are quadratic fits of $A_{\text{overlap}}$ ($A'_{\text{overlap}}$).

Figure 4(b) shows the spatial dependence of $A'_{\text{overlap}}$ for Manhattan JJs with $W_{\text{b}} = W_{\text{t}} = 200$ nm and $\delta W_{\text{offset}} = 25$ nm on a 100-mm diameter wafer.

We can further expand the model by approximating the contribution of sidewalls to $A'_{\text{overlap}}$. The spatially-dependent actual bottom electrode thickness is

$$T'_{\text{b}}(\vec{r}) = T_{\text{b}} \frac{(D' - R)^2 D}{|\vec{r} - \vec{C}|^3}, \tag{3}$$

where $T_{\text{b}} = 35$ nm is the calibrated thickness at $\vec{O}$ under normal incidence ($\alpha = 0$). Approximating the sidewalls as vertical,

$$A'_{\text{overlap}}(\vec{r}) = (W'_{\text{b}}(x) + 2T'_{\text{b}}(\vec{r})) W'_{\text{t}}(y), \tag{4}$$

The modified spatial dependence is shown in figure 4(c). Note that we do not model the effect of shadowing by the bottom electrode during evaporation of the top electrode, which most likely reduces the overlap at the eastern sidewall (evident in figure S8).

Finally, we can model some predictable effects of the first evaporation (for the bottom electrode) on the top electrode. The first evaporation deposits an Al layer above the top resist, effectively increasing its height by $\delta H(\vec{r})$ (also given by the right-hand side of equation 3). More importantly, it also deposits a lip on the southern resist edge for the top electrode (see figure S1), with width $W_{\text{lip}}$ and height $H_{\text{lip}}$:

$$W_{\text{lip}}(\vec{r}) = -T_{\text{b}} \frac{(D' - R)^2 (D'\sin(\alpha) - y)}{|\vec{r} - \vec{C}|^3}, \tag{5}$$

$$H_{\text{lip}}(\vec{r}) = \frac{DW_{\text{t}}}{D'\sin(\alpha) - y}. \tag{6}$$

The shadowing effect of these features makes

$$W'_{\text{t}}(\vec{r}) \approx W_{\text{t}} + \delta W_{\text{offset}} - \begin{cases} W_{\text{lip}}(\vec{r}) + H'(\vec{r})\frac{|y|}{D}, & \text{for } y \geq 0, \\ \max\left(H'(\vec{r})\frac{|y|}{D}, W_{\text{lip}}(\vec{r}) + H'_{\text{lip}}(\vec{r})\frac{|y|}{D}\right), & \text{for } y < 0, \end{cases} \tag{7}$$

where $H'(\vec{r}) = H + \delta H(\vec{r})$ and $H'_{\text{lip}}(\vec{r}) = H_{\text{lip}}(\vec{r}) + \delta H(\vec{r})$. Including all modelled effects leads to $A'_{\text{overlap}}(\vec{r})$ as shown in figure 4(d).

The geometric model predicts that junction conductivity erroneously computed as $G/\Sigma A_{\text{overlap}}$ will show a centre-to-edge decrease. Experimental results for the three Planar $35 \times 35$ wafers clearly show this trend (figures 5(g–i)). In turn, the model predicts that conductivity computed as $G/\Sigma A'_{\text{overlap}}$ will be flat. Due to the inaccuracy of approximating $A'_{\text{overlap}}$ using top-view SEM images, a slight centre-to-edge increase could even be observed. Conductivity computed as $G/\Sigma A'_{\text{overlap}}$ is very uniform for the all-Al wafer but not for the wafers with NbTiN and TiN probing pads, in which a strong centre-to-edge decrease persists. This suggests the possibility of a non-negligible series resistance from the small contact region (nominally $32.4 \times 10^{-2}$ μm²) between each Al electrode and the NbTiN or TiN bays. These contact regions are also susceptible to spatial variation from the shadowing effect. It remains interesting for future work to increase the area of the contact region and to also add a bandaging layer post junction deposition [34, 31] to test if conductivity computed as $G/\Sigma A'_{\text{overlap}}$ can flatten further.

## 6. Conclusions

Table 1 summarizes the findings of our investigation of Dolan and Manhattan JJs on planar and TSV-integrated substrates, spanning yield, conductance CV and frequency RSD at wafer level, as well as average die-level RSD. For planar substrates, Dolan JJs perform best in all categories. In TSV-integrated substrates, Dolan JJs show a marked increase in disorder and decrease in yield, most likely due to their higher susceptibility to resist-height variation. Manhattan JJs are thus the preferred choice for TSV-integrated substrates, but their uniformity must be further improved. First, we must pre-compensate the spatial variation of junction overlap area that arises from the shadowing effect captured by the geometric model. Next, the contribution of contact resistance between the Al electrodes and the NbTiN bays must be quantified and possibly diminished using bandaging layers. These improvements will allow to quantify the intrinsic disorder of Manhattan JJs and approach the $\sim 50$ MHz target that will secure SQP yield by post-fabrication trimming using laser annealing.

## 7. Acknowledgements

We thank Bas van Asten for experimental assistance and David Michalak for valuable discussions. This research is supported by Intel Corporation and by the Office of the Director of National Intelligence (ODNI), Intelligence Advanced Research Projects Activity (IARPA), via the U.S. Army Research Office Grant No. W911NF-16-1-0071. The views and conclusions contained herein are those of the authors and should not be interpreted as necessarily representing the official policies or endorsements, either expressed or implied, of the ODNI, IARPA, or the U.S. Government.

## 8. Additional information

The data shown in all figures of the main text and Supplementary Information are available at `http://github.com/DiCarloLab-Delft/Wafer_Scale_Fab_Data`. Correspondence and requests for additional materials should be addressed to L.D.C. (l.dicarlo@tudelft.nl).

# Supplementary Information

**S1. Fabrication of base layer**

*S1.1. Planar wafers*

We use ⟨100⟩ high-resistivity ($\rho \geq 10$ kΩcm) Si wafers, single-side (double-side) polished for planar (TSV) wafers. Prior to metallization, wafers are cleaned using acetone and isopropanol (IPA). Native oxides are stripped off using a buffered-oxide-etch (BOE 7:1) dip (120 s) and coated with hexamethyldisilazane vapour (HMDS) at 150°C. Next, a niobium titanium nitride (NbTiN) film (200 nm thick) is immediately deposited on the wafers by reactive magnetron sputtering using a 3-inch, 70:30 wt% NbTi target. The film sheet resistance measured using the 4-point method is in the range $1.05 - 1.15$ Ω/□ and the decrease in film thickness from centre to edge is $\sim 25$ nm for the 3 NbTiN-coated wafers. Alternatively, in place of NbTiN, a 160 nm-thick TiN layer is deposited by plasma-enhanced atomic layer deposition (ALD) using 1900 cycles of tetrakis(dimethylamino)titanium (TDMAT) precursor at 300°C. The measured sheet resistance of the TiN film is 1.36 Ω/□. To define the probing pads, a layer of hydrogen silsesquioxane (HSQ) is spun and baked at 300°C, which serves as an inorganic hard mask for wet etching of NbTiN or TiN. The probing pads are defined on the metal-HSQ stack by e-beam lithography using AR-P 6200.13 positive tone resist (1 μm thick). After development with pentyl acetate for 60 s and rinsing with IPA, the metal-HSQ stack is first etched by reactive-ion etching using $SF_6$ and $O_2$. A terminal wet etch to remove any residual metal is done using a 1:1:5 mixture of $H_2O_2$, $NH_4OH$ and $H_2O$ at 35°C. For the bare Si wafer, the Al probing pads are defined along with the junctions in a single writing step.

*S1.2. Fabrication of through-silicon vias*

On NbTiN-HSQ coated wafers, vias of diameter 160 and 400 μm are patterned by photolithography using MX5050 dry-film photoresist and etched using deep-reactive-ion etching (DRIE). The photoresist is stripped using PRS-3000 at 90°C multiple times to completely remove residues. Post TSV integration, the wafers are cleaved at the edges resulting in a 70 mm × 70 mm square wafer. For an actual SQP, this step would be followed by metallization of the vias with TiN deposited by ALD. However, we skip this step in this work. To facilitate uniform dispensing of e-beam resist, a custom chuck is used to hold the square TSV-integrated wafer in place during spinning (in place of standard vacuum chucking). The JJ test structures are then defined on the upper half of the wafer by aligning it with respect to the DRIE layer.

**S2. Fabrication of Josephson junctions**

The fabrication process for each JJ variant is kept constant for all planar and TSV-integrated substrates. A bilayer resist stack is used for both JJ types. For Dolan (Manhattan) junctions, a 400 (200) nm thick layer of poly(methylglutarimide) (PMGI) SF7 is first spun as the support layer, followed by 200 (600) nm poly(methylmethacrylate) (PMMA) 950K A3 (A6) as the imaging layer. Each resist layer is baked for 5 min at 175°C. The JJs are patterned by e-beam lithography using a dose 1850 µCcm$^{-2}$ and step size 4 nm. The PMMA layer is first developed by a 1:3 methyl isobutyl ketone (MIBK) and IPA solution for 60 s, rinsed with additional IPA and dried with N$_2$ gas. The wafer is rinsed in deionized (DI) H$_2$O for 20 s both before and after development of the PMGI layer with MF-321. The JJ patterns are subjected to O$_2$ plasma de-scumming by reactive-ion etching at 10 µbar and 20 W for 50 s followed by oxide strip using a BOE dip for 30 s. The wafers are then immediately transferred to the process chamber of the Al evaporator ensuing pumping in the load lock to $2 \times 10^{-7}$ mbar. For Dolan JJs, we use sample tilt $\alpha = 15°$ and azimuth angles $(\phi_1, \phi_2) = (0°, 180°)$. For Manhattan JJs, we use $\alpha = 35°$ and $(\phi_1, \phi_2) = (0°, 90°)$. For Dolan (Manhattan) JJs, we use 30 (35) nm bottom-electrode thickness and 60 (75) nm top-electrode thickness, deposited at 2 Ås$^{-1}$. The tunnel barrier is grown using static oxidation with 6N purity O$_2$ at 1.3 mbar for 11 min. Following the deposition of the top electrode, terminal capping oxidation is performed using the same oxidation conditions, adding a passivation layer around the junctions.

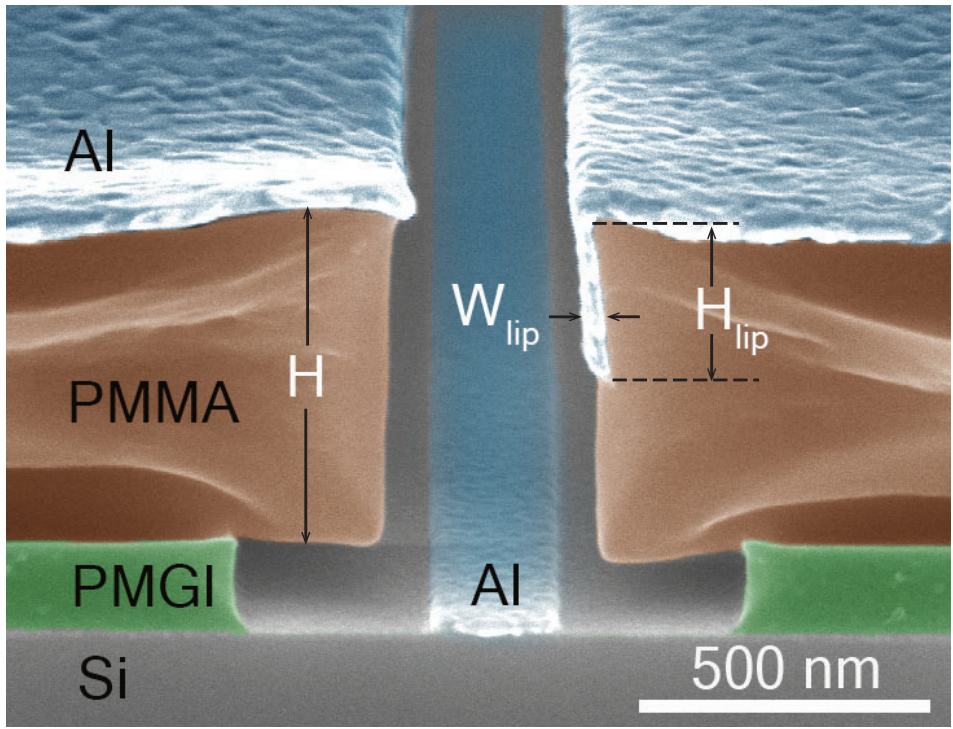

Figure S1: Cross-sectional SEM image of the resist stack of Manhattan JJs, with added false colour to highlight different materials. The cut shown corresponds to the horizontal electrode of a Manhattan JJ near wafer centre. The wafer is cleaved after deposition and lift-off of 20 nm of Al for both bottom and top electrodes at $\alpha = 35$ °. The Al thickness is intentionally reduced to minimize buckling of the resist stack. This image is taken at 76° tilt, using low beam current (10 µA) and accelerating voltage (5 kV) to minimize distortion of the resist stack. The large undercut for the PMGI layer is created by the higher dissolution rate of PMGI during development using MF-321, which is based on tetramethyl ammonium hydroxide.

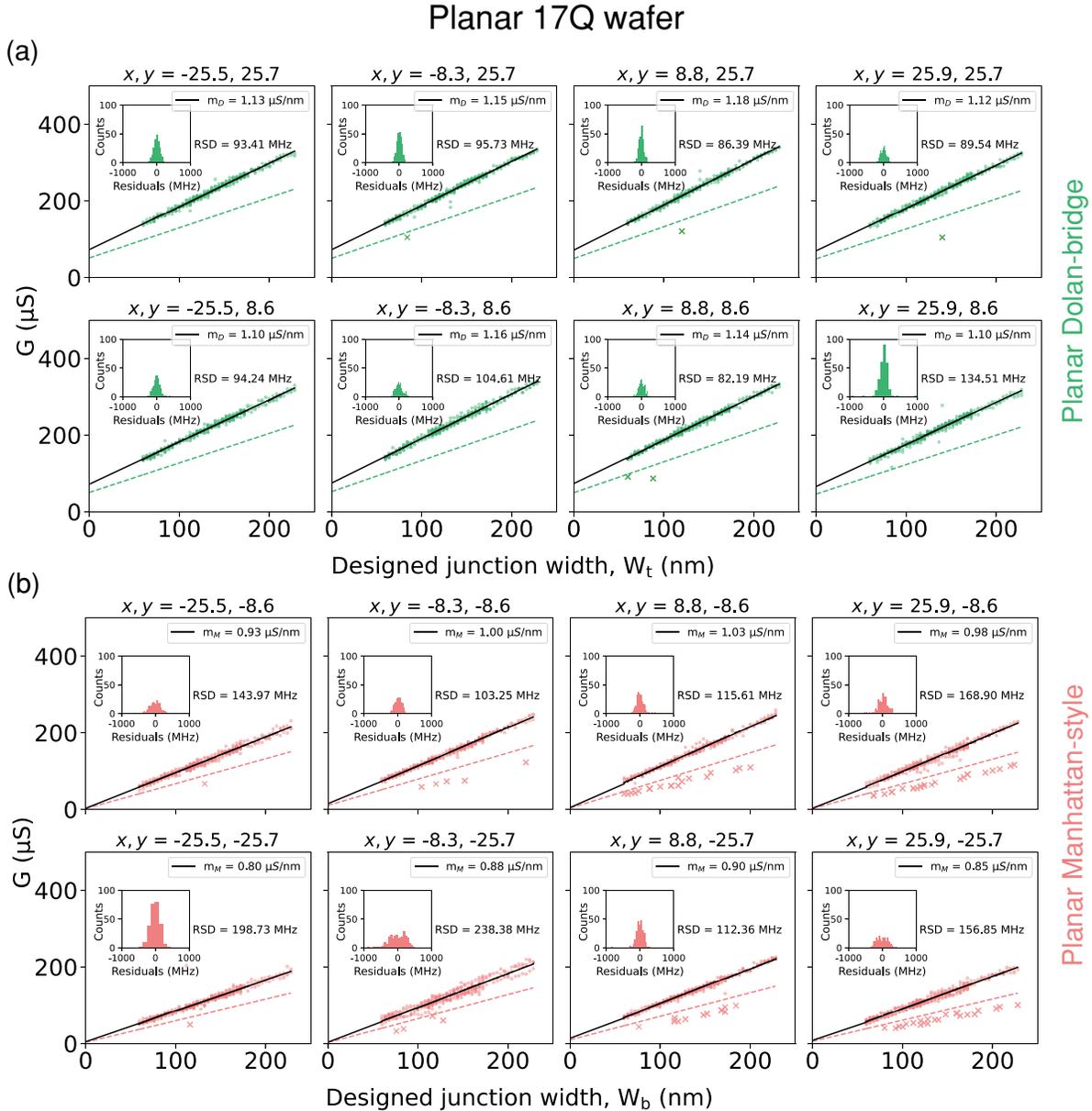

Figure S2: Die-level linear regression analysis of conductance as function of designed junction width of the variable electrode of (a) Dolan and (b) Manhattan JJ pairs on the Planar 17Q wafer. Coordinates shown at the top of each panel indicate die center (relative to wafer centre). The dotted coloured lines correspond to the best fit of the initial regression. The crosses below the dotted coloured line are interpreted as half-open junctions. The filtered data are then fit again with a second regression. Insets: The residuals from the second regression fit are plotted in units of predicted qubit frequency. The frequency RSDs obtained from these histograms are plotted in figures 2(c) and 3(c) of the main text. We note that the sequence of electrode deposition for Dolan junctions in the Planar 17Q wafer was accidentally reversed during fabrication. A significant sidewall contribution to the conductance may explain the $\sim 70$ µS intercept in this dataset.

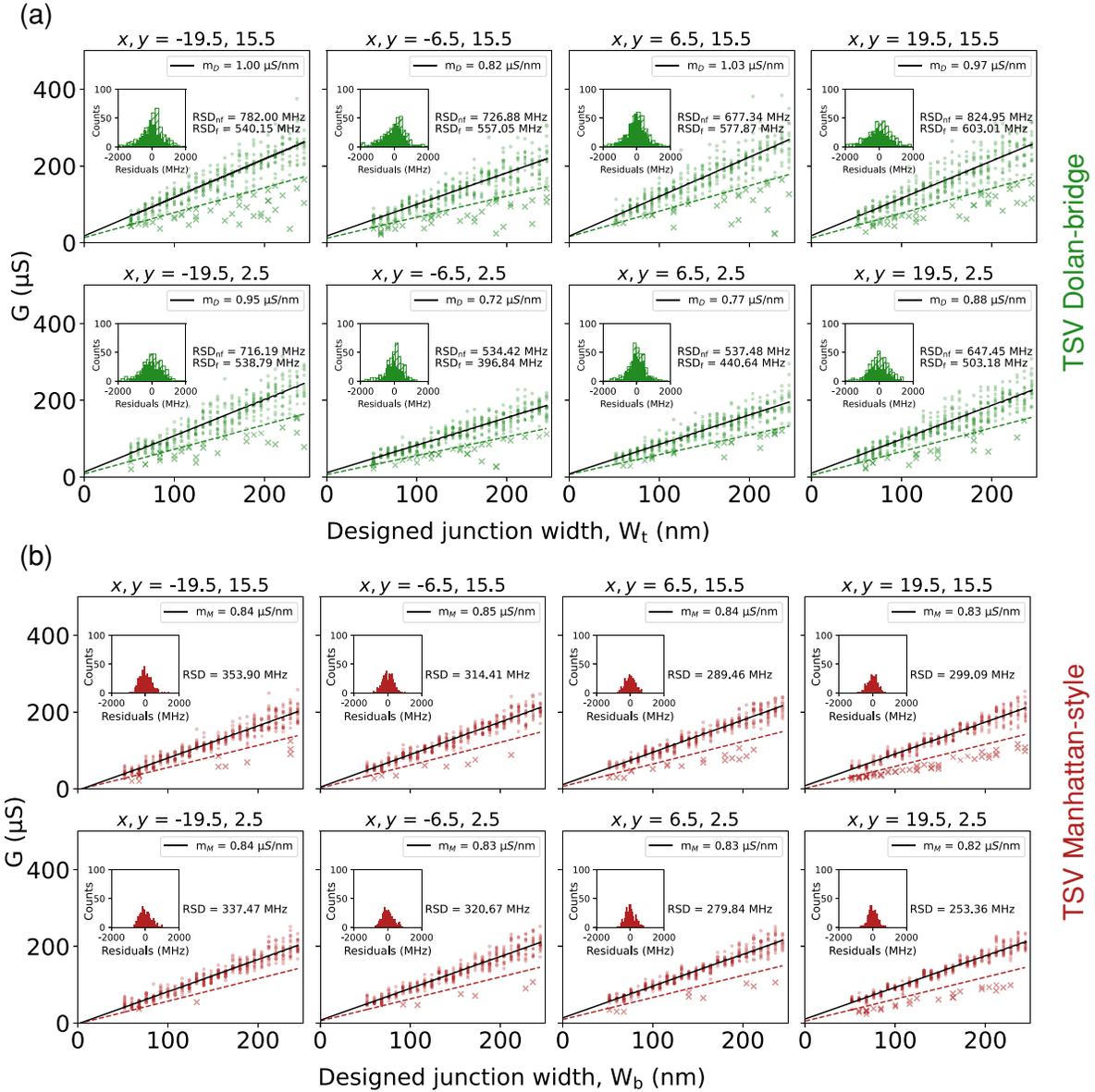

Figure S3: Die-level linear regression analysis of conductance as a function of designed width of the variable electrode of (a) Dolan and (b) Manhattan JJ pairs on the TSV 17Q wafer. Coordinates shown at the top of each panel indicate die center (relative to wafer centre). The outliers are filtered using the method described in figure S2. Due to the large spread in conductance in the TSV Dolan data, the initial regression filter does not clearly discriminate half-open junctions. The die-level RSD is therefore calculated from unfiltered ($RSD_{nf}$) as well as filtered ($RSD_f$) conductance data.

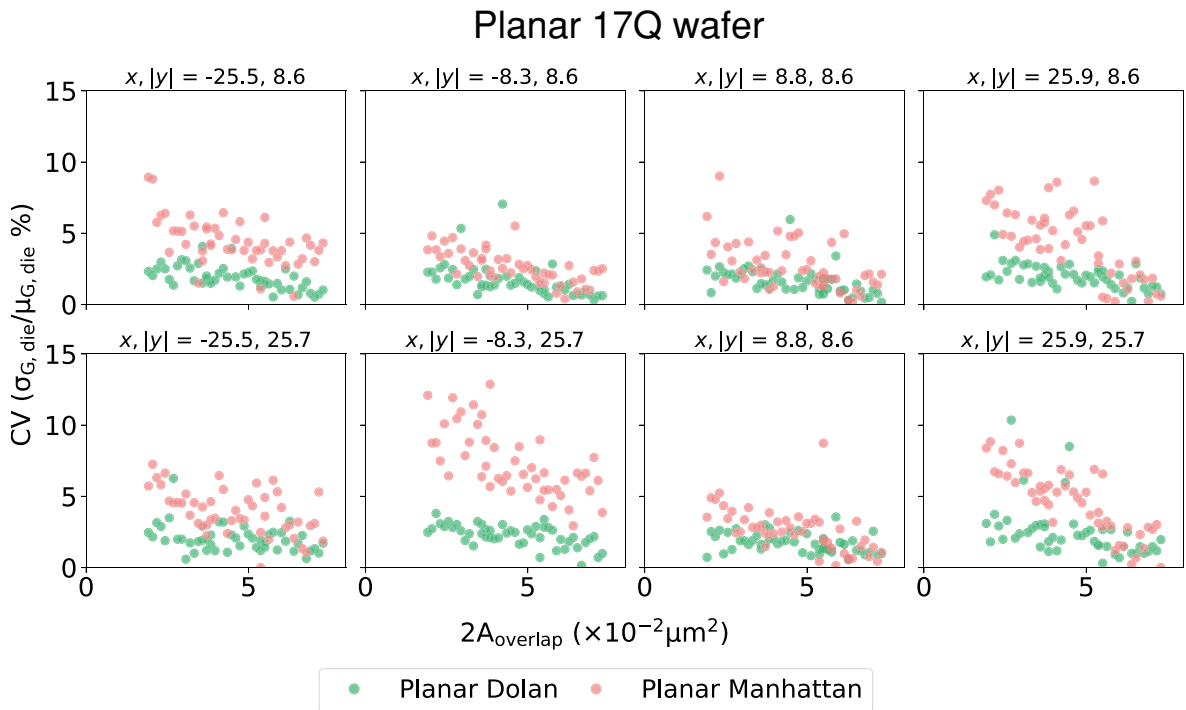

Figure S4: Die-level conductance CV as a function of the designed overlap area for Dolan (green) and Manhattan (red) JJ pairs fabricated on the Planar 17Q wafer. Coordinates shown at the top of each panel indicate die center (relative to wafer centre).

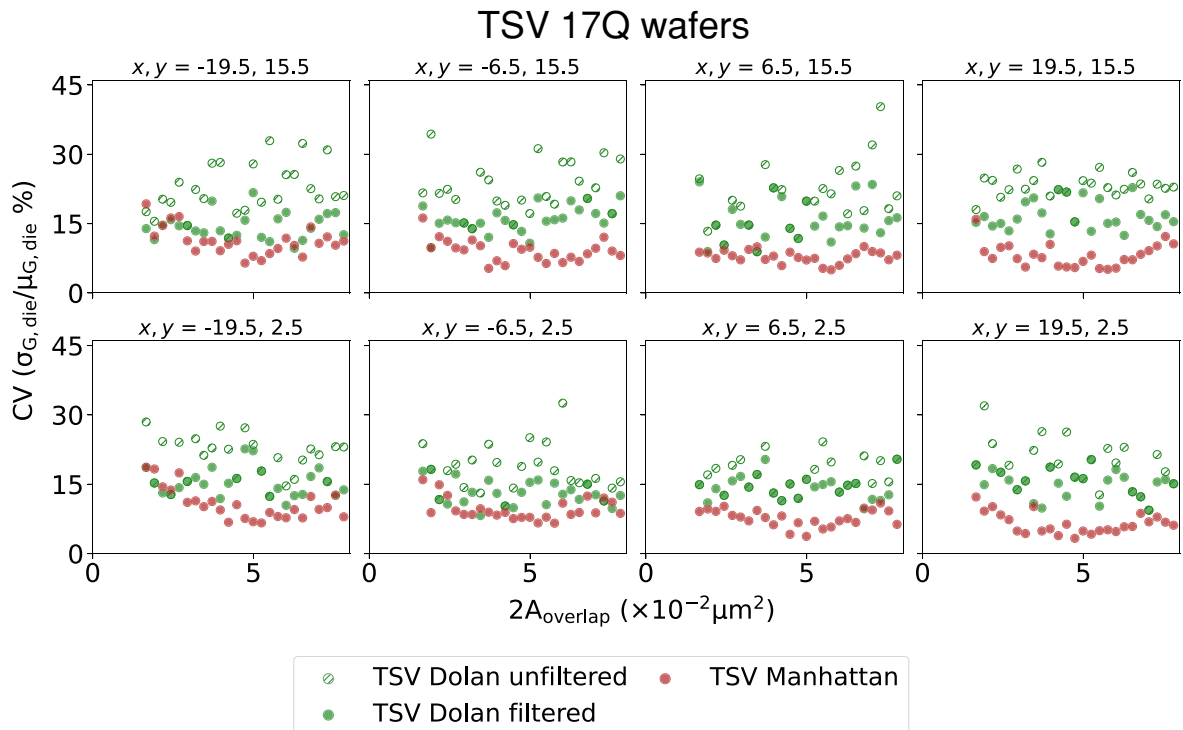

Figure S5: Die-level conductance CV as a function of the designed overlap area for unfiltered (hatched) and filtered (green) Dolan and for Manhattan (red) JJ pairs fabricated on the TSV 17Q wafer. Coordinates shown at the top of each panel indicate die center (relative to wafer centre).

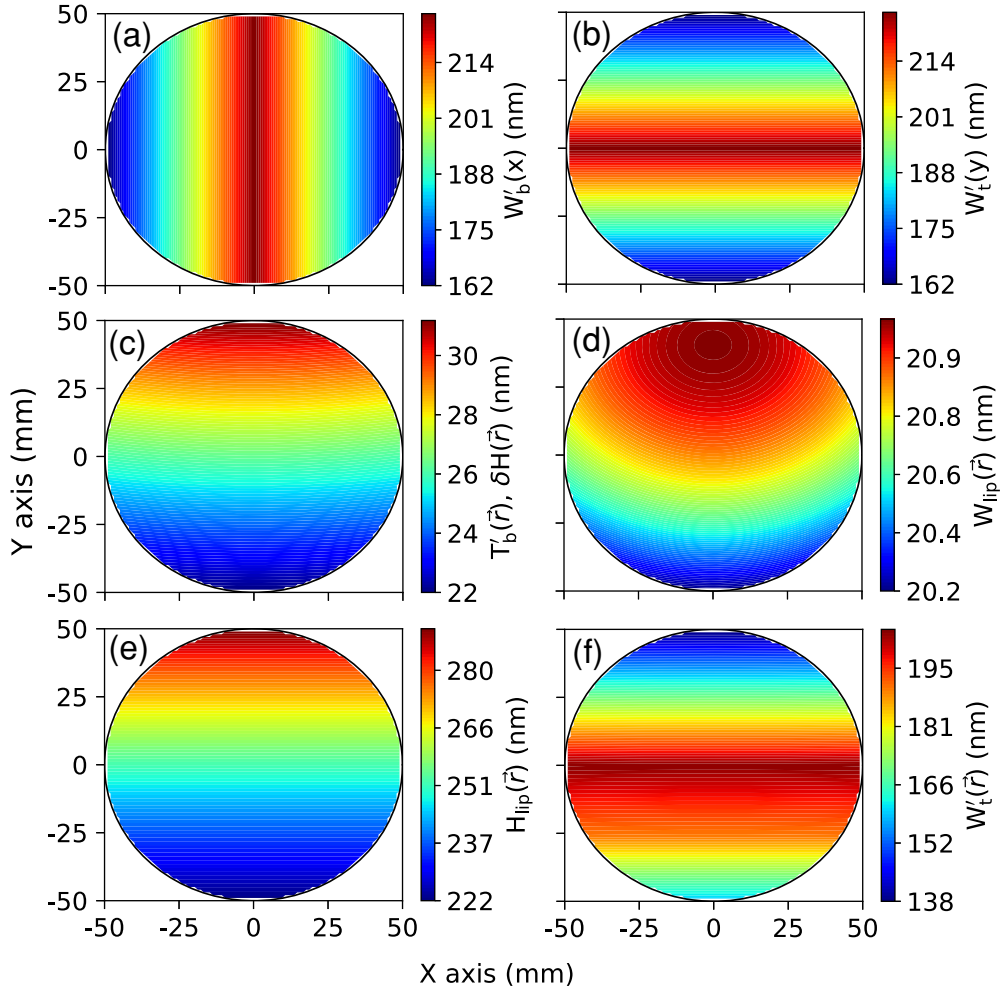

Figure S6: Contour plots of spatially dependent variables of the geometric model. (a) Actual bottom electrode width $W'_b(x)$ as per equation 2. (b) Actual top electrode width $W'_t(y)$ as per equation 2. (c) Actual bottom electrode width $T'_b(\vec{r})$ as per equation 3. (d) Lip width $W_{lip}(\vec{r})$ as per equation 5 (see figure S1 for reference). (e) Lip width $H_{lip}(\vec{r})$ as per equation 6 (see figure S1 for reference). (f) Actual top electrode width $W'_t(\vec{r})$ as per equation 7.

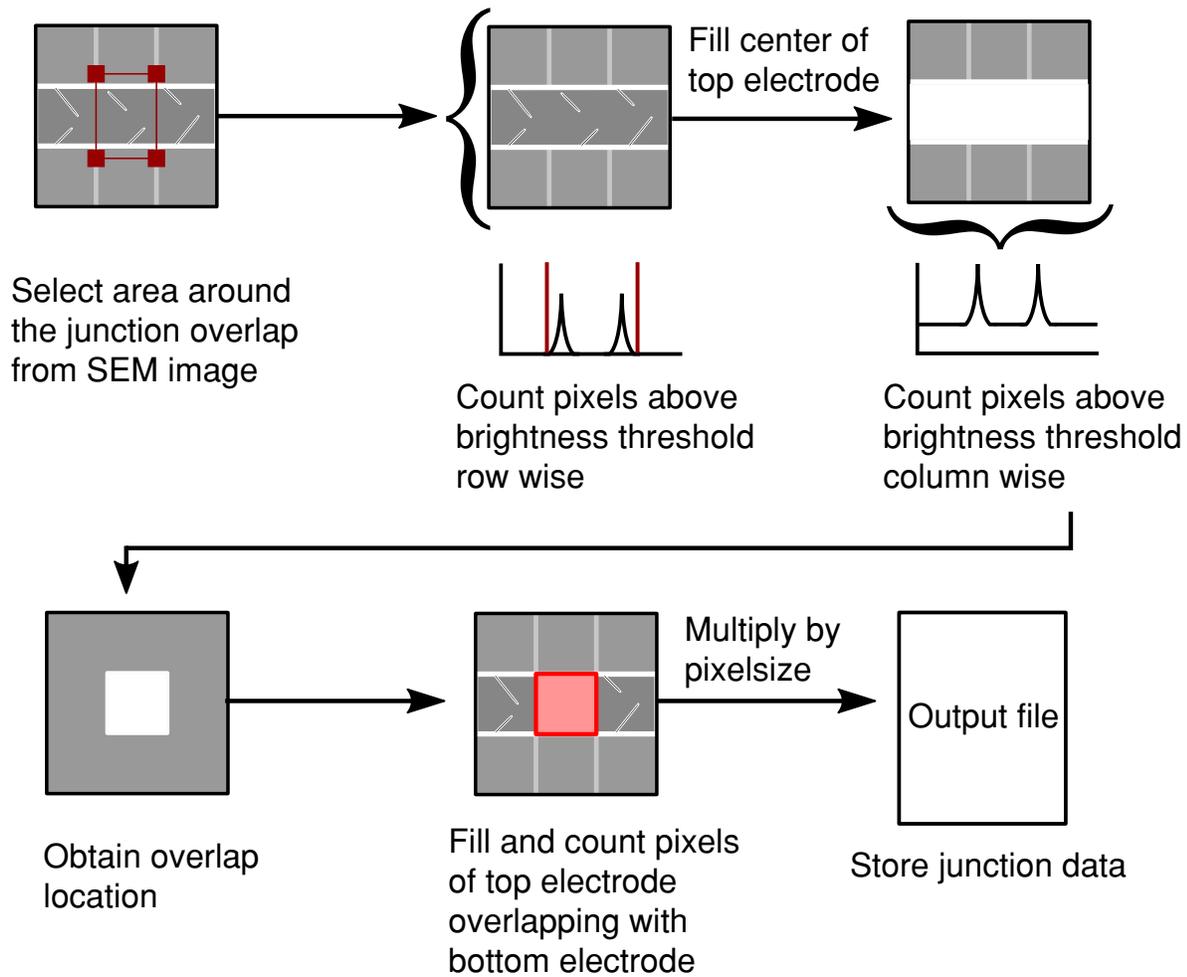

Figure S7: Process flow of our Python-based image analysis software analyzing SEM images of Manhattan junctions. The average pixel value of the image is first calculated, then multiplied by a set range of thresholds between 1.0 and 2.0 to detect the edge. The top electrode is filled first to reduce pixel noise before detecting the edges of the bottom electrode. By summing the pixels row (column) wise above the threshold, the centre row (column) corresponding to the edge is obtained. The actual width $W'_\text{t}$ ($W'_\text{b}$) is calculated from the mean of the non-zero distance between the edges for each threshold. The defined range of thresholds is used to filter the image and obtain the best edge. The overlap area $A'_\text{overlap}$ is obtained by summing the pixels between the outer row edges of $W'_\text{t}$ over the inner column edges of $W'_\text{b}$.

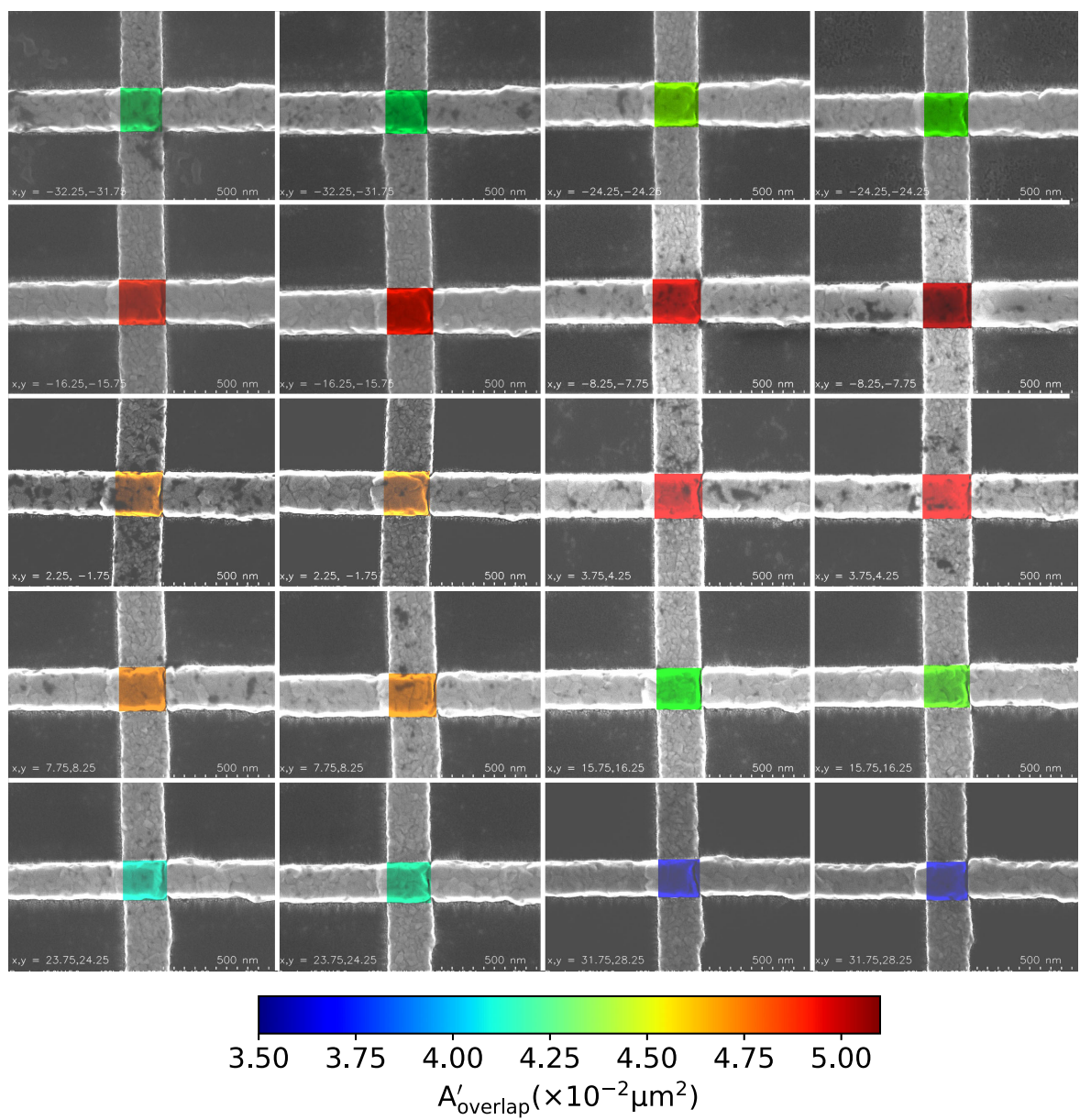

Figure S8: Compilation of a subset of SEM micrographs used to extract actual electrode widths and junction overlap areas for several of the Manhattan-junction pairs fabricated on the planar TiN wafer. The images are acquired from test pads positioned diagonally across the wafer shown by the black dotted line in figure 5(b). Images of junctions for a JJ pair are placed side by side. Each image is labelled with the coordinates of the JJ pair relative to wafer centre.